\documentclass[preprint]{aastex}

\usepackage{color}
\usepackage{amsmath}

\title{Superorbital Phase-Resolved Analysis of SMC X-1}

\author{Chin-Ping Hu, Yi Chou, Ting-Chang Yang, Yi-Hao Su}
\affil{Graduate Institute of Astronomy, National Central University, Jhongli 32001, Taiwan}
\email{Hu: m929011@astro.ncu.edu.tw, Chou: yichou@astro.ncu.edu.tw}

\begin{document}

\begin{abstract}
The high-mass X-ray binary SMC X-1 is an eclipsing binary with an orbital period of 3.89 d. This system exhibits a superorbital modulation with a period varying between $\sim$40 d and $\sim$65 d. The instantaneous frequency and the corresponding phase of the superorbital modulation can be obtained by a recently developed time-frequency analysis technique, the Hilbert-Huang transform (HHT). We present a phase-resolved analysis of both the spectra and the orbital profiles with the superorbital phase derived from the HHT. The X-ray spectra observed by the Proportional Counter Array onboard the { \it Rossi X-ray Timing Explorer} are fitted well by a blackbody plus a Comptonized component. The plasma optical depth, which is a good indicator of the distribution of material along the line of sight, is significantly anti-correlated with the flux detected at $2.5-25$ keV.  However, the relationship between the plasma optical depth and the equivalent width of the iron line is not monotonic: there is no significant correlation for fluxes higher than $\sim35$ mCrab but clear positive correlation when the intensity is lower than $\sim20$ mCrab.  This indicates that the iron line production is dominated by different regions of this binary system in different superorbital phases.  To study the dependence of the orbital profile on the superorbital phase, we obtained the eclipse profiles by folding the All Sky Monitor light curve with the orbital period for different superorbital states. A dip feature, similar to the pre-eclipse dip in Her X-1, lying at orbital phase $\sim0.6-0.85$, was discovered during the superorbital transition state. This indicates that the accretion disk has a bulge that absorbs considerable X-ray emission in the stream-disk interaction region.  The dip width is  anti-correlated with the flux, and this relation can be interpreted by the precessing tilted accretion disk scenario.
\end{abstract}


\section{Introduction}
SMC X-1, first discovered by \citet{Leong1971}, is an eclipsing high-mass X-ray binary (HMXB) consisting of a neutron star with a mass of 1.06 $M_{\odot}$ \citep{vandermeer2007} and a B0 I supergiant companion, Sk 160, with a mass of 17.2 $M_{\odot}$ \citep{Reynolds1993}. This system exhibits X-ray pulsation with a 0.71 s period \citep{Lucke1976}, an orbital eclipse every 3.89 d \citep{Schreier1972}, and superorbital modulation (\citealt{Gruber1984}) with a period varying between $\sim40$ d and $\sim65$ d \citep{Trowbridge2007, Hu2011}. The superorbital modulation is interpreted as an obscuring effect caused by a precessing warped and tilted accretion disk \citep{Wojdowski1998}.  After the launch of the All Sky Monitor (ASM) onboard the { \it Rossi X-ray Timing Explorer (RXTE)}, the variation of the superorbital modulation period of SMC X-1 was studied by various time-frequency techniques, e.g., the Morlet wavelet transform \citep{Ribo2001}, the dynamic power spectrum \citep{Clarkson2003}, the slide Lomb-Scargle periodogram \citep{Trowbridge2007}, and the Hilbert-Huang transform \citep{Hu2011}. Among these techniques, the HHT, proposed by \citet{Huang1998}, can provide a well-defined instantaneous frequency, as well as the corresponding phase, of the superorbital modulation.

The X-ray spectra are good indicators of the properties of the central X-ray source and the environment of the binary system. \citet{Woo1995} studied the variation in the spectral properties for a complete orbital cycle and identified the wind dynamics in this binary system using the data collected by {\it Ginga}. \citet{Wojdowski2000} analyzed the X-ray spectrum observed by the {\it Advanced Satellite for Cosmology and Astrophysics} during an eclipse and found that it is inconsistent with the line-driven wind model proposed by \citet{Blondin1995}. \citet{Vrtilek2005} reported the results of spectral analysis of eight {\it Chandra} observations, which covered different orbital phases and superorbital states.  They found that the spectra in the superorbital high state are independent of the orbital phase, whereas the low-state spectra depend strongly on the orbital phase.

Combining timing and spectral analysis could facilitate to further study of the nature of this binary system. For example, \citet{Naik2004} applied energy-resolved timing analysis to the {\it BeppoSAX} observations and found that the pulse profiles of the soft thermal component and the hard power-law component are different. \citet{Hickox2005} made a similar analysis and presented pulse phase-resolved spectral analysis of SMC X-1 using {\it Chandra} and {\it XMM-Newton} observations. The variations in the pulse profiles of different spectral models were attributed to a twisted, pulsar-illuminated inner accretion disk. Because the {\it Chandra} and {\it XMM-Newton} observations covered parts of the superorbital phase, the long-term variation in the pulse profile can roughly support the disk and beam geometry.

If the number of observations is large enough to cover most of the superorbital modulation, it is possible to further study the variation in the spectral properties versus the superorbital modulation in detail with phase-resolved spectral analysis. For example, \citet{Leahy2001} studied the spectral variation of the 35 d superorbital modulation of Her X-1 using {\it Ginga} observations.  \citet{Naik2003} analyzed the variations in the spectral parameters, especially the iron line intensity and equivalent width (EW), in different superorbital states of LMC X-4 and Her X-1.  Because the X-ray observations of SMC X-1 made by the {\it RXTE} cover most of the superorbital phase, the only difficulty in performing phase-resolved spectral analysis is that the period of the superorbital modulation is not stable, so the corresponding phase is hard to define. However, this difficulty can be solved by using an advanced time-frequency analysis method, the HHT, which can provide a well-defined phase function for modulation with a variable period.

The orbital profile, in addition to the spectral behavior, was found to vary with the superorbital modulation. \citet{Trowbridge2007} studied the orbital profile of different superorbital phases, but they had insufficient statistics to demonstrate a relationship between the orbital profile and superorbital phase.  The increased number of RXTE/ASM observations since then, combined with the superorbital phase defined by the HHT, should provide a statistically sound basis for studying orbital profile variations.

We present our studies of the superorbital phase-resolved analysis of the X-ray spectra and the variation in the orbital profile in this paper. In section \ref{obs}, we briefly introduce the observations made by the Proportional Counter Array (PCA) onboard the {\it RXTE} and the light curve collected by the ASM. The spectral models, the variation in the spectral parameters versus the superorbital phase, and the variation in the orbital profile with the superorbital phase, are described in Section \ref{result}. Finally, we discuss our results in section \ref{discussion}, in particular the variation in the iron line EW and the mechanism that causes the variation in the X-ray dip.

\section{Observation and Data Analysis}\label{obs}

\subsection{RXTE PCA}
The PCA onboard the {\it RXTE} consists of five proportional counter units (PCUs) with an energy range of 2--60 keV. The field of view is limited by the collimator to $\sim 1^\circ$. SMC X-1 was observed frequently by the PCA from 1996 August 28 to 2004 January 30.  red Between 1997 and 1998, the PCA made $1-3$ observations on SMC X-1 per month.  These observations provided $\sim 40$ data ponits randomly distributed over the superorbital cycle.  Other 5 series of consecutive observations made in 1996, 2000, 2003, and 2004 covered $\lesssim 0.5$ superorbital cycles.  All the data are archived on the website of the High Energy Astrophysics Science Archive Research Center (HEASARC) of the National Aeronautics and Space Administration (NASA).  The spectra were analyzed  using all of the Standard-2 mode data except for those between 1998 October 16 and 1998 December 4, which were contaminated by an outburst from the nearby X-ray pulsar XTE J0111.2-7317 \citep{Chakrabarty1998}.  Furthermore, two PCA observations (OBSID 30125-05-01-00 and 30125-05-03-02) were rejected because they contained insufficient photons to yield meaningful spectral fittings. 

The Standard-2 mode data provide 129 energy channels with an energy range of 2 to 60 keV and 16 s timing resolution; however, only the energy channels  between $\sim$2.5 and 25 keV were used for spectral fitting. For the low-energy boundary, we ignored the first three channels because the PCA calibration is not accurate for channels $1-3$.  Among the five PCUs, we extracted only the spectra from PCU2 because it was always in operation. The spectra, response matrices, and background models were created using the {\tt FTOOLS} analysis software. Furthermore, a 1\% systematic error was added to the error of each spectrum as in the manipulation by \citet{Inam2010}, which was suggested by \citet{Wilms1999}.

The time span covers August 1996 to January 2004, which corresponds to {\it RXTE} gain epochs 3 to 5. To correct the problem with gain-drift of {\it RXTE}, we normalized the flux of individual observations to the nearest epoch Crab observation. 

\subsection{RXTE ASM}
Although the PCA observations cover most of the superorbital states, they are still insufficient for studying the relationship between the orbital and superorbital profiles. Thus, we used ASM data to investigate the variation in the orbital profiles. The ASM onboard the {\it RXTE} continuously swept the entire sky once every 90 min for the entire {\it RXTE} lifetime. Its energy range is 1.3 to 12.1 keV. The summed band dwell light curve collected since MJD 50,134 was analyzed for orbital profile variations.  Because the ASM gain of ASM has changed moderately in the last two years \citep{Levine2011}, the light curve is slightly noisy after MJD 55,200 and even the superorbital modulation cannot be recognized after MJD 55,600.  Thus, the data collected after MJD 52,000 were excluded from this study.

\section{Results}\label{result}
\subsection{High and Low State X-ray Spectra}
To obtain the variation in the spectral parameters with respect to the superorbital phase, we first removed the data within orbital phase of 0.87 to 1.13 according to the orbital ephemeris proposed by \citet{Wojdowski1998} to avoid variations caused by eclipses.  We then fitted the combined spectra for both the high and low state with different models in order to select the model that best describes the PCA spectra.  The superorbital phases from \citet{Hu2011} were adopted to define four states according to the superorbital profile: the high state (0.19 -- 0.54), the low state (0.73 -- 1.05), and two transitions: the ascending state (0.05 -- 0.19) and descending state (0.54 -- 0.73).  In previous studies, we noted that the flux in the same state may differ with the superorbital cycle; e.g., the high state count rate of the 40-d superorbital cycle is generally lower than that of the 65-d cycle because the modulation amplitude and period are anti-correlated \citep{Hu2011}.  Because the number of X-ray photons is sufficiently high in the high state, it is improper to combine all the high state observations from different superorbital cycles, which may have different spectral properties.  Thus, we combined only the spectra in a series of consecutive observations made in December 2003 to obtain the spectral model of the high state. For the low state, combining all the uneclipsed observations is acceptable because the count rate is relatively low.  We did not combine the spectra in the transition phases because the spectral properties change dramatically during these states. The spectra, together with the corresponding background and RMF files, were combined using the {\tt addspec} command. 

The {\tt XSPEC v12.8.0} package of {\tt HEAsoft} was applied for spectral fitting.  We tried two spectral models to describe the spectra of both the high and low states. Model 1 contains a blackbody component plus a simple power law with a high-energy cutoff \citep{Woo1995, Naik2004, Vrtilek2005, Hickox2005}.  Model 2 contains a blackbody component plus a Comptonized component \citep{Naik2004, Vrtilek2005}, which describes the Comptonization of soft blackbody photons in a hot plasma \citep{Titarchuk1994}.  A Gaussian line with a central energy of 6.4 keV, which corresponds to the central energy of the K$\alpha$ emission line of iron atoms, was added during spectral fitting.  As a result, Model 1 can be described as follows: 
\begin{equation}
I(E)=\exp[-n_H\sigma(E)]\times\left[f_{bb}(E)+f_{pl}(E)f_{cut}(E)+f_{Fe}(E)\right]
\end{equation}
where $\sigma(E)$ is the photoelectric cross section, $n_H$ is the equivalent hydrogen column density, $f_{bb}(E)$ is the blackbody emission, $f_{pl}(E)$ is the power law model, $f_{cut}(E)$ is a multiplicative model that represents a high-energy exponential cutoff, and $f_{Fe}(E)$ is a Gaussian iron emission line.  Model 2 can be described as follows: 
\begin{equation}
I(E)=\exp[-n_H\sigma(E)]\times\left[f_{bb}(E)+f_{comp}(E)+f_{Fe}(E)\right]
\end{equation}
where $f_{comp}(E)$ is the Comptonized component, which was implemented as compTT in the XSPEC package, and the input soft photon energy is set to the same value as the blackbody temperature. 

\begin{figure}
\plotone{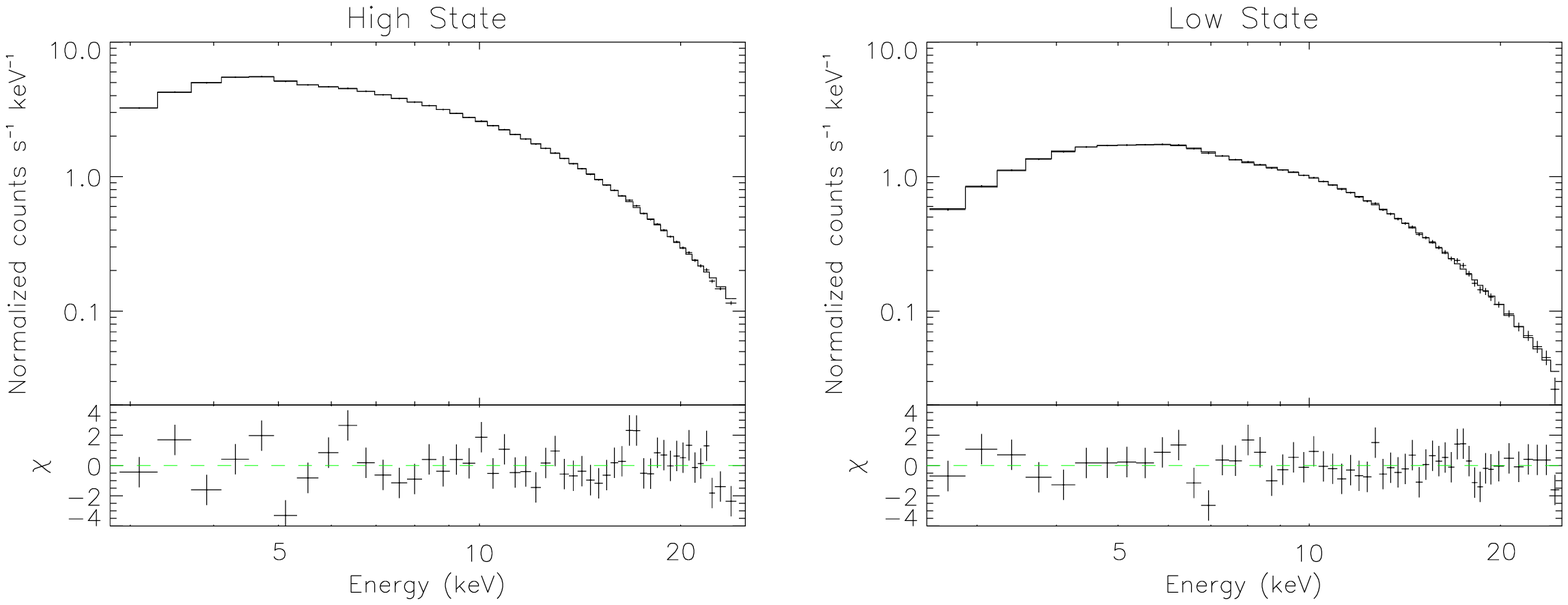} \caption{{\it RXTE} PCA spectra obtained in both superorbital high state (left) and low state (right).  The histograms are the best-fit models, and the residuals in $\chi$ are shown in the corresponding lower panel. \label{example_spectra}}
\end{figure}

The best-fit parameters are shown in Table \ref{fitting_diff_model}.  The reduced $\chi^2$ revealed that using the inverse Comptonized model provided significantly better fitting than using the power law with a high-energy cutoff.  Thus, we selected Model 2 to describe the spectra from all the individual PCA observations in further analysis.  The folded count rate spectra, corresponding models, and residuals are shown in Figure \ref{example_spectra}.  From the observed parameters, we found that both $n_H$ and the plasma optical depth ($\tau$) in the low state are significantly higher than those in the high state.  In addition, the EW of iron line was also calculated by using the {\tt eqwidth} command in {\tt XSPEC}.  The plasma temperature ($kT_e$), line width ($\sigma_{line}$), and EW are lower in the low state than in the high state.

\subsection{Superorbital Phase-Resolved Variation in Spectral Parameters}
\begin{figure}
\plotone{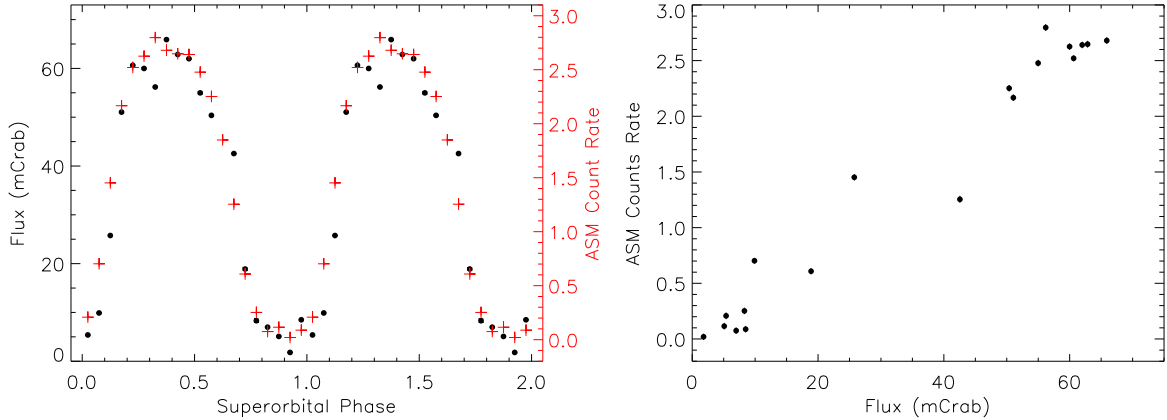} \caption{Left: Phase-resolved flux obtained from spectral fitting of PCA data (black dots) and the corresponding ASM count rate (red crosses).  Right: Correlation between PCA flux and ASM count rate. \label{pca_asm}}
\end{figure}

Before the phase-resolved spectral analysis, we examined the correlation between the PCA observations and the {\it RXTE}/ASM data on the basis of the phase obtained from the HHT.   We divided the superorbital phase into 20 bins and only one of them contained no PCA observation.  The mean flux of individual bins obtained by spectral fitting of PCA observations, together with the folded ASM light curve with the same 20 bins, are shown in Figure \ref{pca_asm}.  The linear correlation coefficient between the PCA flux and the corresponding ASM count rate is 0.98 with a null hypothesis probability of $3.7\times 10^{-7}$, which indicates a very strong correlation. This correlation means that the PCA observations can almost reproduce the ASM ones according to the definition of the superorbital phase by the HHT. 

\begin{figure}
\center
\plotone{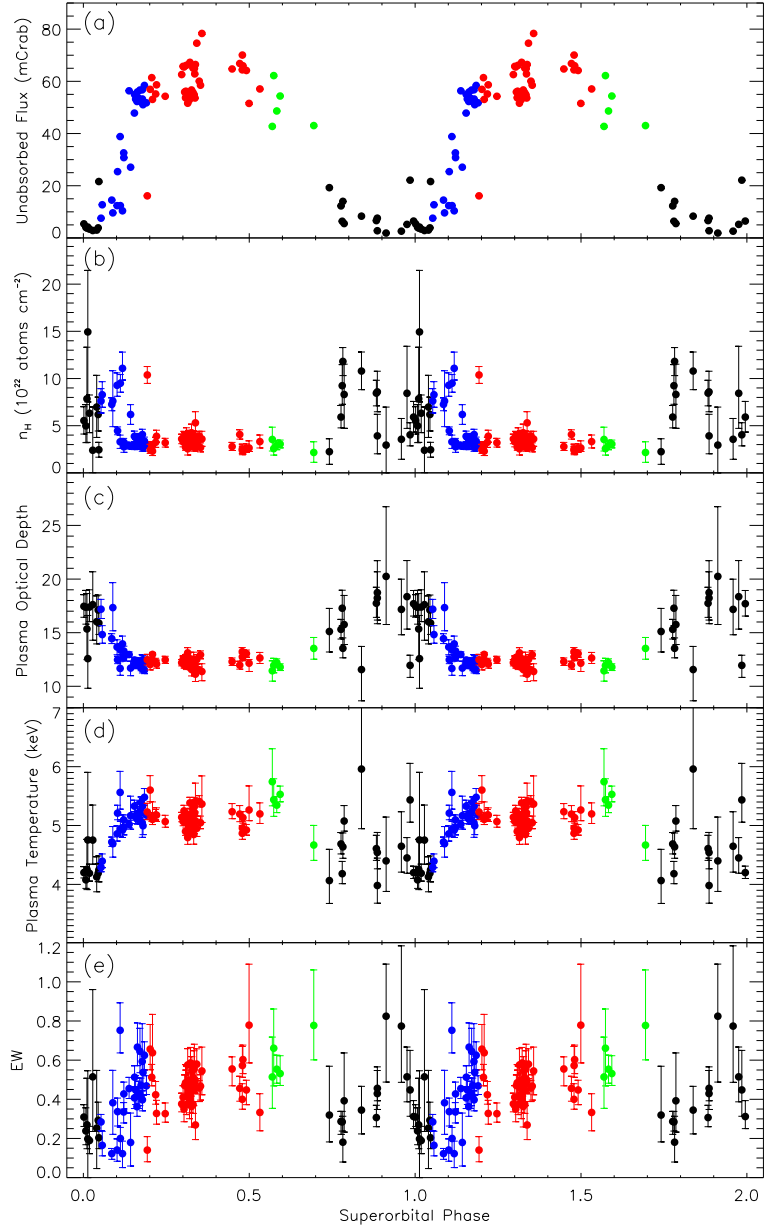} \caption{Spectral fitting parameters folded with superorbital period.  (a)--(e) Variations in the unabsorbed flux, hydrogen column density ($n_H$), plasma optical depth ($\tau$), plasma temperature ($kT_e$), and equivalent width (EW) of iron line, respectively. Black, blue, red, and green symbols represent the spectral parameters in the low, ascending, high, and descending states, respectively.  The typical errors in the unabsorbed flux are $\lesssim 1$ mCrab, which is extremely small to be displayed in the figure.    \label{parameter_superorbital} }
\end{figure}

To obtain the variation in the spectral parameters that could reveal the emission and absorption properties of the different phases of disk precession, we folded all the spectral parameters according to the superorbital phase.  Figure \ref{parameter_superorbital} shows the variations in the unabsorbed flux, $n_H$, $\tau$, $kT_e$, and the EW of the iron line in different superorbital phases.  The unabsorbed flux, which is normalized by the nearest epoch Crab observation, shows similar properties in the binned PCA observations and folded ASM light curve, e.g., an asymmetric superorbital profile.  In the high state, the $n_H$ values are consistent with those in \citet{Inam2010} and the $\tau$ values also fall in a narrow range.  We further examined the correlations and found that both $n_H$ and $\tau$ are anti-correlated with the flux.  The linear correlation coefficient between $n_H$ and the flux is $-0.69$ with a null hypothesis probability of $5.4 \times 10^{-16}$, whereas the linear correlation coefficient between $\tau$ and the flux is $-0.81$, with a null hypothesis probability of $5.8 \times 10^{-22}$.  Although both $n_H$ and $\tau$ show strong anti-correlations, $n_H$ is strongly influenced by the soft X-ray band, to which the PCA is insensitive.  Furthermore, the $n_H$ values show great diversity during the low and ascending states.  In contrast, the $\tau$ values show less diversity and have a stronger anti-correlation with the flux.  Thus, we chose $\tau$ as the indicator of the material in the line of sight.  The variation in $\tau$ and $n_H$ in the descending state seems to differ from that in the ascending state.  This may represent either sampling bias caused by insufficient statistics or an indication of different absorption properties on the ascending/descending sides of the warp region.  On the other hand, the linear correlation coefficient between $kT_e$ and the flux is $0.66$ with a null hypothesis probability of $3.9 \times 10^{-15}$, which indicates a strong positive correlation.  This is not surprising because the reprocessing region is farther from the central X-ray source in the low state than in the high state. 

\begin{figure}
\center
\plotone{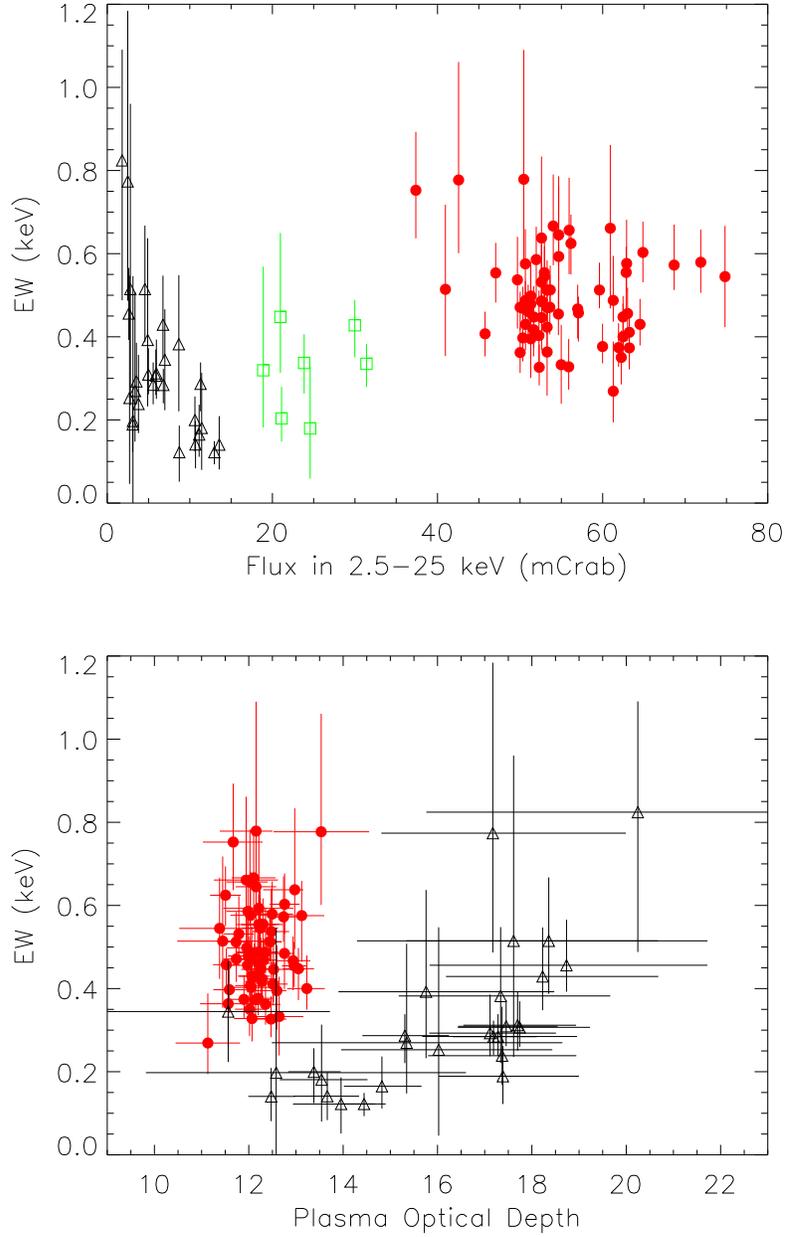} \caption{Upper panel: Relationship between the EW of the iron line and the observed flux.  Red filled circles denote data points with fluxes higher than 35 mCrab; black open triangles denote data points lower than 19 mCrab.  Green open squares are those data in transition between the high- and low-intensity regions.  Lower panel: Relationship between the EW of the iron line and $\tau$.  Data in transitions are omitted.  \label{corr_ew_tau} }
\end{figure}

Another interesting parameter that is related to the flux is the EW of the iron line.  The correlation between the EW and the flux is relatively complex, as shown in the upper panel of Figure \ref{corr_ew_tau}.  We found that the correlations in the high-intensity and low-intensity regions are probably different.  The linear correlation coefficient between the EW and the flux for those data points with fluxes higher than 35 mCrab is  $-0.19$ with a null hypothesis probability of $0.06$, which indicates a marginal anti-correlation.  However, when the flux is lower than 19 mCrab, the linear correlation coefficient between the EW and the flux is $-0.59$ with a null hypothesis probability of $1.9\times 10^{-4}$, which indicates a much stronger anti-correlation.  The relation between the EW and $\tau$ is plotted in the lower panel of Figure \ref{corr_ew_tau}. The distributions of those two data groups are obviously distinct, and the correlations are also different significantly.  The linear correlation coefficient between the EW and $\tau$ at fluxes higher than 35 mCrab is $0.13$ with a null hypothesis probability of 0.22, which indicates no significant correlation. On the other hand, the linear correlation coefficient between the EW and $\tau$ at fluxes lower than 19 mCrab is 0.64 with a null hypothesis probability of $5.3\times 10^{-5}$, which shows a strong positive correlation.  The different correlations between the EW and $\tau$ indicate different origins of the iron line production, which will be discussed in section \ref{discussion}. 

\subsection{Superorbital Phase-Resolved Variation in Orbital Profile}
In addition to the spectral behavior, the variation in the orbital profile with the superorbital phase is also an interesting issue for further study.  Because the sampling of PCA observations is insufficient to investigate variations in the orbital profile, we used the data collected by the ASM to conduct this portion of our study.  Each superorbital cycle was first equally divided into 20 subsets according to the superorbital phase. All the subsets of the same superorbital phase were then folded with the orbital ephemeris provided by \citet{Wojdowski1998}.  We therefore obtained 20 superorbital phase-resolved eclipsing profiles, as shown in Figure \ref{fold_lc_all} (a).  The eclipse profiles for the ascending, high, descending, and low states are shown in Figure \ref{fold_lc_all} (b) -- (e), respectively.  The profile of the high state resembles that of a typical total eclipsing X-ray binary with a sharp eclipse, whereas those of the ascending, descending, and low states show greater variation in the uneclipsed region.  For the ascending and descending states, we eliminated the data points near the high state within $\sim0.05$ superorbital cycles so that the most interesting characteristic of the orbital profile, the dip-like feature at orbital phase $\sim0.6-0.85$, would be more visible.  A broad dip feature between orbital phases 0.5 and 0.85 can also be seen in the ascending state, whereas a narrower dip feature during orbital phases 0.65 and 0.85 can be observed in the descending state.  In the low state, the dip feature is unclear owing to low photon statistics.  The dip feature is believed to represent absorption by the bulge in the accretion stream-disk interaction region. 

Because the variation in the orbital profile is strongly related to the superorbital phase, a two-dimensional folded light curve is a good way to investigate the relationship between the orbital and superorbital profiles, as suggested by \citet{Trowbridge2007}.  We first defined a data window in the superorbital phase domain with a size of 0.05 cycles and folded the data points in the window with the orbital ephemeris. The window was then moved forward by a step of 0.01 cycles to obtain the next orbital profile. This process was repeated until the end of the data set. Finally, all the profiles were combined into a three-dimensional map, as shown in Figure \ref{dynamic_fold}. The uneclipsed count rates of all the orbital profiles were normalized to 1, and the resulting map was smoothed by a Gaussian filter to enhance the eclipse and dip features.

From the dynamic folded light curve, we found that a sharp eclipse feature can be seen throughout both transition states and the high state ($\sim0.05-0.75$). In addition, a major dip appears in the descending and early low state.  It is centered at orbital phase $\sim0.7$ and increases in width as the flux decreases.  The dip feature in the ascending state is less visible and wider than that in the descending state, and the relationship between the dip width and flux is harder to obtain.  In the deep low state, the eclipse and dip features cannot be recognized because of limited ASM sensitivity.

\section{Discussion and Conclusions}\label{discussion}
SMC X-1 exhibits an obvious superorbital modulation, the period of which changes dramatically with time.  The { \it RXTE}/PCA observations covered most of the superorbital phases, so the data provided fruitful information on the spectral properties of different superorbital states.  \citet{Inam2010} analyzed all the spectra observed by the PCA. All the $n_H$ values, including those in different orbital and superorbital states, were found to increase as the X-ray flux decreased.  This may be due to absorption by the companion or a warped region in the accretion disk. However, further details of the relationship between the spectral indices and superorbital phases were still unknown. The most difficult challenge, the definition of superorbital phases of variable periodicity, is solved by using the HHT.  This research demonstrated phase-resolved spectral analysis of the uneclipsed observations.  First, we found that the combined spectra are better described by the inverse Comptonized component than by a power-law with a high-energy cutoff.  In addition, both $n_H$ and $\tau$ show an anti-correlation with the flux, but the correlation between $\tau$ and the flux is more significant than that between $n_H$ and the flux.  Thus, we chose $\tau$ as an indicator of absorption by the material along the line of sight in the PCA energy range.  

Comparing $n_H$ and $\tau$, we found that the EW of the iron line exhibits a more complex relationship with the flux. The correlation between the EW and the flux when the flux is lower than 19 mCrab shows similar behavior to that of LMC X-4 obtained by \citet{Naik2003}, who explained the variation in the EW by the presence of two producing regions.  In the high-intensity state, the iron line is dominated by emissions near the central region of the compact object, and the EW is almost constant.  When the disk precesses to the low-intensity state, the central region is almost obscured by the inner warp, and the iron line is dominated by another region far from the compact object.  Unfortunately, the variation in $n_H$ with the superorbital phase of LMC X-4 is not available.  In SMC X-1, we use another indicator, $\tau$, to represent the materials that reprocessed the X-rays.  From the relations between the EW and $\tau$ we found a positive correlation when the flux was lower than 19 mCrab, which contains the low state, the early ascending state, and possibly the late descending state, although no samples were available.  This indicates that SMC X-1 also contains a weaker iron-line emitting region far from the central neutron star.  In the high-intensity state, the EW of SMC X-1, like that of LMC X-4, remains in a relatively stable region and does not show a significant correlation with the flux and $\tau$.  TSMC X-1 differs from LMC X-4 in that the EWs of SMC X-1 in the high-intensity region are larger.  The EW values of SMC X-1 when the flux is larger than 35 mCrab are similar to that of Her X-1 obtained by \citet{Leahy2001}, who studied the variations in the iron line EW during the 35-d superorbital cycle of Her X-1 using  {\it Ginga} observations.  The EW of Her X-1 in the main high state has a mean value of 0.48 keV and a standard deviation of 0.12 keV. In our case, those EWs at fluxes higher than 35 mCrab have a mean value of 0.49 keV and a standard deviation of 0.11 keV, which is consistent with that of Her X-1.  Although the flux varies dramatically from 70 mCrab to 35 mCrab, the obscuring effect rather than the absorption effect dominates the variation in the flux: thus, neither $\tau$ nor the EW  is strongly correlated with the flux. However, when the flux drops to less than $\sim20$ mCrab, the iron line production is dominated by another region more distant than the inner warp of the accretion disk.  At the same time, absorption by materials far from the central region dominates the flux variation.  Thus, both $\tau$ and the EW show strong anti-correlations with the flux.  Using {\it Chandra} observations, \citet{Vrtilek2005} shows that the iron line of SMC X-1 consists of at least two components: a 6.4 keV K$\alpha$ line superposing on a broad Fe line.  The Fe line in this study is a combination of those components due to limited spectral resolution of {\it RXTE}, and the variation of EW may indicates varying contributions of them.  Thus, we could not identify how the individual component varies. The variation of all the line components on both the superorbital and orbital phases can be achieved after the X-ray observatories with high spectral resolutions, like {\it Chandra} and {\it XMM-Newton}, make enough amount of observations. 

\begin{figure}
\plotone{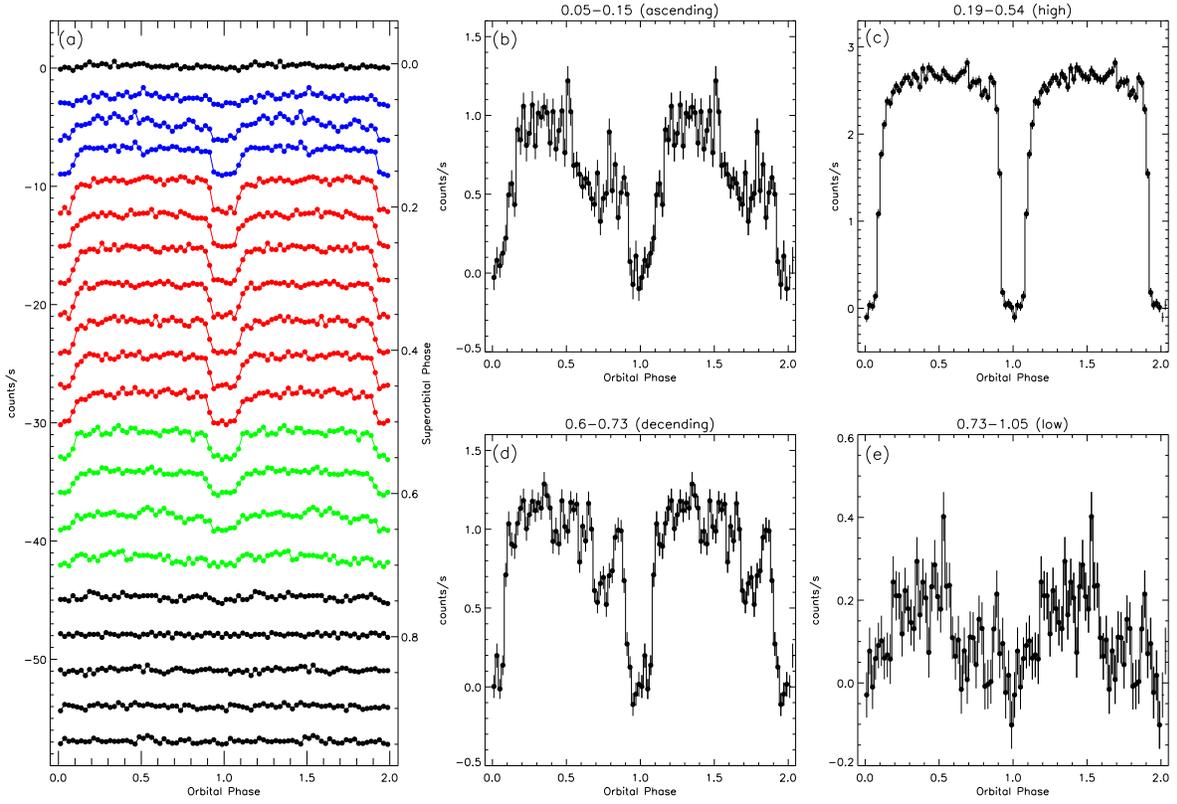} \caption{(a) Folded light curves of all the superorbital phases, each vertically shifted by $-3$ for ease of viewing. Different colors represent the orbital profile in different superorbital states, as defined in Fig \ref{parameter_superorbital}. (b) -- (e) Four  folded light curves of different superorbital states (ascending, high, descending, and low, respectively). \label{fold_lc_all}}
\end{figure}

\begin{figure}
\plotone{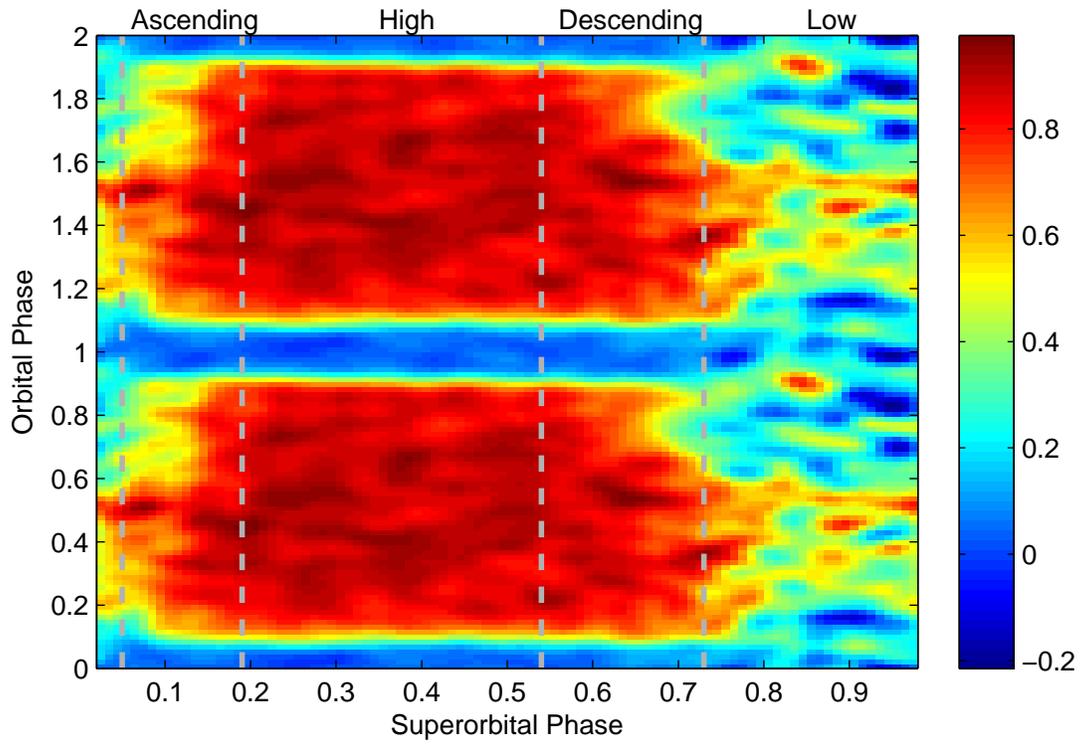} \caption{Dynamic folded light curve of SMC X-1. Color denotes the normalized count rate; gray dashed lines are the boundaries of different superorbital states.  The dip feature can be easily obtained from this figure. \label{dynamic_fold}}
\end{figure}

From the variation in the orbital profile (Figures \ref{fold_lc_all} and \ref{dynamic_fold}), we found an absorption dip at orbital phase $\sim 0.6-0.85$, the width of which increases as the count rate decreases in the transition and low states.  Dips in X-ray binary systems are believed to be caused by absorption of the central X-ray emission in the impact region of the accretion stream and disk. Although SMC X-1 is an HMXB system, the steady high X-ray intensity could indicate the existence of a stream-fed accretion disk \citep{Woo1995}. In the superorbital high state, the inclination angle of the tilted disk is low, and we can observe the central X-ray source directly during the uneclipsed phase. As the disk precesses to the transition state, the inclination angle becomes higher; the central X-ray source begins to be gradually obscured by the inner warped region of the accretion disk, and the X-ray intensity start to decrease. At the same time, the bulge in the outer rim of the accretion disk is also lifted, becoming closer to our line of sight. Because the bulge is co-rotating with the binary system, we see the periodic absorption dip at orbital phase $\sim0.6-0.85$.  

\citet{Woo1995} studied the light curve and spectral variations in 1.3 orbital cycles of {\it Ginga} observations. The extended and asymmetric eclipse transitions agree with the line-driven stellar wind model proposed by \citet{Blondin1995}, although the distribution of circumstellar material was further modified by \citet{Wojdowski2000} and \citet{Wojdowski2008}. However, we found that the orbital profile varies greatly with the superorbital phase. Thus, the variation in the orbital profile in our analysis is more likely related to the precession of the accretion disk than to the distribution of circumstellar material. \citet{Woo1995} suggested that the dip feature is caused by the absorption of the accretion stream on the basis of the observed variation in $n_H$ in orbital phase 0.9 but this feature could not be directly obtained in their light curve (see Figure 2 in \citet{Woo1995}). Instead, a tiny dip can be marginally obtained in orbital phase $\sim0.7$. We could not identify the superorbital state of the {\it Ginga} observation because it occurred before the launch of the {\it RXTE}, but it is unlikely to be the low state owing to the high count rate. If the {\it Ginga} observation was made during the superorbital high state, it implies that the narrow, shallow dip can be observed even in the high state. High-state dips could not be observed in the ASM light curve, probably because of the limited sensitivity. We look forward to the data collected by the {\it Monitor of All-sky X-ray Image} ({\it MAXI}), which has better sensitivity, for verification.

\citet{Moon2003} and \citet{Trowbridge2007} also mentioned the dip feature of SMC X-1 and associated it with the light curve dips of Her X-1, although no further studies of the SMC X-1 dip were made. The dips of Her X-1 were first discovered by \citet{Giacconi1973}, and they can be further divided into two groups, pre-eclipse dips and anomalous dips \citep{Moon2001}. A series of extensive studies have been made since then, e.g., \citet{Shakura1998, Moon2001, Igna2011} and references therein. The dips obtained in orbital phase $\sim0.6-0.85$ of SMC X-1 may be associated with the pre-eclipse dip distributed in orbital phase $0.7-0.9$ of Her X-1. Interestingly, the pre-eclipse dip of Her X-1 would migrate toward earlier orbital phases when the disk precesses. We did not detect this migration behavior but obtained the variation in the dip width of SMC X-1.  Thus, the mechanism of the dip of SMC X-1 is probably not as complex as that of Her X-1. Future extensive studies with observations of higher sensitivity could unveil the detailed properties of the dips of SMC X-1.

\acknowledgments
This research made use of the {\it RXTE}/PCA data provided by the High Energy Astrophysics Science Archive Research Center of NASA's Goddard Space Flight Center. The data collected by the ASM are provided by the ASM/{\it RXTE} teams at MIT and at the {\it RXTE} SOF and GOF at NASA's GSFC. This research was supported by grant NSC 100-2119-M-008-025 and NSC 101-2112-M-008-010 from the National Science Council of Taiwan.

\epsscale{0.7}
\epsscale{1.0}

\epsscale{0.7}
\epsscale{1.0}

\clearpage

\begin{table}
\begin{tabular}{c|cc|cc}
\hline 
State & \multicolumn{2}{c|}{High} & \multicolumn{2}{c}{Low}\tabularnewline
\hline 
Model & Model 1 & Model 2 & Model 1 & Model 2\tabularnewline
\hline
\hline 
$n_{H}$ & $4.15\pm0.15$ & $3.90\pm0.16$ & $6.61_{-0.64}^{+0.71}$ & $7.32_{-0.61}^{+0.64}$\tabularnewline
$\Gamma$ & $1.47\pm0.01$ & -- & $1.13\pm0.04$ & --\tabularnewline
$E_{c}$ (keV) & $13.87\pm0.15$ & -- & $11.68_{-0.25}^{+0.34}$ & --\tabularnewline
$E_{f}$ (keV)& $20.73\pm0.57$ & -- & $16.01_{-0.86}^{+0.79}$ & --\tabularnewline
$kT_e$ (keV)& -- & $5.18\pm0.04$ & -- & $4.61\pm0.06$\tabularnewline
$\tau$ & -- & $12.34\pm0.11$ & -- & $15.23\pm0.42$\tabularnewline
$\sigma_{line}$ (keV) & $2.17\pm0.06$ & $1.65\pm0.06$ & $0.74_{-0.1}^{+0.11}$ & $0.54_{-0.12}^{+0.10}$\tabularnewline
EW (keV) & $0.59\pm0.2$ & $0.37^{+0.02}_{-0.01}$ & $0.34_{-0.05}^{+0.04}$ & $0.24\pm0.03$\tabularnewline
$\chi_{\nu}^{2}$ (dof) & 3.83 (41) & 1.71 (42) & 1.51 (46) & 0.84 (47)\tabularnewline
\hline
\end{tabular}

\caption{Best-fit parameters for superorbital high and low states. Unit of $n_{H}$ is $10^{22}$atoms cm$^{-2}$. $kT_e$ is the plasma temperature, $\tau$ is the plasma optical depth, and $\chi_{\nu}^{2}$ is the reduced $\chi^2$ with degree of freedom (dof). }\label{fitting_diff_model}
\end{table}

\end{document}